\documentstyle[12pt,titlepage]{article}
\newcommand{\be}{\begin{equation}}
\newcommand{\ee}{\end{equation}}
\newcommand{\br}{\begin{eqnarray}}
\newcommand{\er}{\end{eqnarray}}

\newcommand{\half}{\frac{1}{2}}
\def\b{{\beta}}
\def\g{{\gamma }}
\def\G{{\Gamma}}

\def\S{{\Sigma}}
\def\l{{\lambda} }

\def\m{{\mu}}
\def\n{{\nu}}

\def\dirac#1{\setbox0=\hbox{$#1$}\rlap{\hbox                   to 
\wd0{$\hss\mkern1mu/\hss$}} \box0 }

\begin{document}

\begin{titlepage}
\begin{center}

{\bf ASYMPTOTIC FREEDOM AND CONFINEMENT IMPLY \\
SPONTANEOUSLY BROKEN CHIRAL SYMMETRY\\
IN QUANTUM CHROMODYNAMICS}\\
\vspace{.5in}

R. Acharya ($\ast $)\\
Physics Department, Arizona State University, Tempe, AZ 85287\\
and\\

P. Narayana Swamy ($\dagger$ )\\
Physics  Department, Southern Illinois University,  Edwardsville, 
IL 62026\\
\end{center}
\vspace{.3in}

\begin{center}

{\bf Abstract}
\end{center}

We utilize Coleman's theo\-rem and show that Quantum 
Chro\-mo\-dynamics based on asymp\-totic freedom and  confinement 
{\it  must\/}  have chiral symmetry realized as  a  spontaneously 
broken symmetry.

\vspace{1.5in}

($\dagger$ ): e-mail address: pswamy@siue.edu\\
($\ast $): e-mail address: acharya@phyast.la.asu.edu

\vfil

\end{titlepage}

In  a  vector-like gauge theory, conventional wisdom  says  that, 
except for the explicitly broken $U_A(1)$ symmetry of the  flavor 
axial-vector  current  of  fermions by instantons  of  the  gauge 
fields,  the  other  global  flavor  chiral  symmetries  will  be 
spontaneously   broken.  Coleman  and  Witten   [\ref{CW}]   have 
established  spontaneously  broken chiral symmetry in  the  large 
$N_c$ limit of Quantum Chromodynamics.  Although a great deal  of 
work   exists  on  this  subject  [\ref{tHooft}]  using   several 
approaches, a general rigorous proof of this however is not known 
.  As pointed out by Weinberg [\ref{QFT}], it can be  shown  that 
``$ SU(3)\times SU(3)$ symmetry for massless $u,d,s$ quarks  must 
in  fact be spontaneously broken in QCD but it is more  difficult 
to  show  on  the  basis of QCD that  the  $SU(2)  \times  SU(2)$ 
symmetry  with only $u,d$ massless quarks is  also  spontaneously 
broken".

As emphasized by Adler [\ref{AdlerRMP}], two of the  relativistic 
field  theory  models which exhibit dynamical  spontaneous  scale 
invariance breaking are: 1) Johnson-Baker-Wiley model [\ref{BJW}] 
of Quantum Electrodynamics in which chiral symmetry is explicitly 
broken  and 2) asymptotically free gauge theories (QCD) in  which 
chiral  symmetry  is spontaneously broken. Thus it is  the  sense 
that the underlying theory exhibiting spontaneously broken  scale 
invariance  cannot  be manifestly chiral  symmetric  (Wigner-Weyl 
mode).   However, two models do not appear to be a  large  enough 
sample for such a generalization although the result itself might 
be suggestive. 

We  shall first establish the following result by the  method  of 
{\it reductio ad absurdum\/}: Confinement and Asymptotic  freedom 
imply that scale invariance {\it must\/} be broken  spontaneously 
in Quantum Chromodynamics (QCD) with quarks and gluons.

Unbroken scale invariance states that 
\be
 Q_D(t) \, |0>=0,
\ee
where
\be
Q_D(t) = \int d^3x \; D_0({\bf x},t)
\ee
is the dilatation charge defined in terms of $D_{\m}({\bf x},t)$, 
the dilatation current.
This is equivalent to
\be
\partial^{\m}D_{\m}\, |0>=0.
\ee
We  may  now invoke Coleman's theorem [\ref{Coleman}],  which  is 
based  on the Federbush-Johnson-Jost-Schroer theorem  [\ref{PCT}, 
\ref{Federbush},  \ref{Strocchi}]. This is valid  for  continuous 
symmetries  and  states that the invariance of the vacuum is  the 
invariance  of  the  world. Consequently the  divergence  of  the 
dilatation current  must vanish identically:
\be
\partial^{\m}D_{\m}=0.
\ee
On  the other hand, the divergence of the dilatation  current  is 
determined by the trace anomaly in QCD [\ref{Collinsetal}]:
\be
\partial^{\m}D_{\m}= \half \frac{\b(g)}{g}G^a_{\m\n}G^{\m\n}_a  + 
\sum_i m_i [1+\g_i(\theta)]{\bar \psi}_i \, \psi_i.
\ee
The second term vanishes since $m_i=0, \; i=u,d,s$ in the  chiral 
limit. Consequently, the beta function must vanish:
\be
\b(g)=0.
\ee
We  observe  that  the Callan-Symanzik beta  function  is   gauge 
independent   in   the   minimal  subtraction   scheme   in   QCD 
[\ref{Yndurian}].  Now  we  can ask where the zero  of  the  beta 
function is: either when $g=g^* \not= 0$ or when $g=0$.  We  know 
that the asymptotically free theory of QCD must have $g=0$ as  an 
ultraviolet  stable fixed point. The beta function which  due  to 
asymptotic  freedom is negative at small coupling (provided  $N_F 
\le 16$), remains negative  for all couplings in such a way  that 
the  effective  coupling constant grows without bounds  at  large 
distances ({\it i.e.,\/} quark confinement) [\ref{Zin}]. In other 
words,  confinement  requires  that the curve  of  $\b(g)$  as  a 
function  of $g$ remains below the positive $g$ axis,   decreases 
as  $g$  increases and never turns  over  [\ref{Marciano}].  This 
rules  out  the  first possibility. As a  consequence,  the  only 
possible solution is
\be
g=0
\ee
and the theory is reduced to a triviality.

We therefore conclude by {\it reductio ad absurdum\/}, that scale 
invariance must be broken spontaneously by the QCD vacuum state: 
\be
Q_D(t) \, |0>\not=0.
\ee
Since QCD is a nontrivial theory with $g \not=0$, scale invariance is also 
explicitly broken by the trace anomaly, as in Eq.(5), As a consequence, there 
is no Goldstone theorem [\ref{GSW}] and there is no dilaton. In other words, 
scale invariance is broken both spontaneously by the vacuum state and 
explicitly by the trace anomaly.

At this point we shall first establish the vanishing  of 
the  commutator  of the dilatation charge $Q_D(t)$and  the  axial 
charge $Q_5^a(t)$ at $t=0$.

We  shall begin with the fact that $Q^a(0), \; a=1, 2,  \cdots  , 
8$, the generators of $SU(3)$ transformations satisfy the algebra
\be
[\, Q^a(0), \, Q^b(0)\, ] = i f_{abc} Q^c(0).
\ee
This can be consistent with 
\be
[\, Q_D(0),\, Q^a(0)\,]= -i d_Q Q^a(0)
\ee
only if [\ref{Callan}]  the scale dimension of charge   is  zero: 
$d_Q=0$ . This can be explicitly proved as follows.

Consider the double commutator 
\be
[    Q_D(0),   [   Q^a(0),   Q^b(0)]   ]=    if_{abc}    [Q_D(0), 
Q^c(0)]=f_{abc}d_Q Q^c(0).
\ee
Employing Jacobi identity, this reduces to
\be
-id_Q [Q^a(0), Q^b(0)] + id_Q [Q^b(0), Q^a(0)]=f_{abc} d_Q Q^c(0)
\ee
which can be satisfied only if $d_Q=0$. We can now extend this to 
the axial-charge $Q^a_5$ by employing the commutators
\be
[Q^a_5(t), Q^b_5(t)]=if_{abc}Q^c(t),
\ee
and
\be
[Q_D(0), Q^a_5(0)]=-id_{Q5}Q^a_5(0),
\ee
where  $d_{Q5}$  is  the scale  dimension  of  the  axial-charge. 
Evaluating  the double commutator of the above with $Q_D(0)$,  we 
find
\be
[Q_D(0),  [  Q^a_5(0), Q^b_5(0)) ] ]=  if_{abc}[Q_D(0),  Q^c(0)]= 
d_{Q} f_{abc}Q^c(0)=0.
\ee
If we now employ Jacobi identity, we obtain
\be
- 2i d_{Q5}[Q^a_5(0), Q^b_5(0)]= 2d_{Q5}f_{abc}Q^c(0)=0.
\ee
Again,  this can be satisfied only if the scale dimension of  the 
axial  charge  vanishes:  $d_{Q5}=0$ and hence  it  follows  that 
$[Q_D(0),  Q^a_5(0)]=0$.  By introducing $e^{iHt}$ on  the  left, 
$e^{-iHt}$ on the right and $1=e^{-iHt}\; e^{iHt}$ in the middle, 
we  further  establish  the vanishing of the  commutator  of  the 
dilatation  charge and the axial charge for any  time:  $[Q_D(t), 
Q^a_5(t)]=0$.

We shall now establish the result that spontaneously broken scale 
invariance implies spontaneously broken chiral symmetry. We shall 
again employ the method of {\it reductio ad absurdum\/}. We  have 
already proved that scale invariance is spontaneously broken  and 
thus  $Q_D(t)|0>\not=0$.  This means that the  dilatation  charge 
does not annihilate the vacuum but instead the operation produces 
another  state: $Q_D(t)|0>=|\phi_1 (t)>$, and $|\phi_1(t)>  \not= 
|0>$. The dilatation charge is time dependent and hence does  not 
commute with the Hamiltonian. As a result, the states produced by 
repeated  application  of  the  dilatation  charge  operator  are 
neither vacuum states nor necessarily degenerate. 
Does  spontaneously broken scale invariance  imply  spontaneously 
broken chiral symmetry? Let us assume to the contrary that chiral 
symmetry  is  unbroken,  $Q_5^a(t)|0>=0$.  Since  $Q_5^a(t)$  and 
$Q_D(t)$ commute, we therefore obtain 
\be Q_5^a(t)Q_D(t)|0>=Q_5^a(t)|\phi_1(t)>= Q_D(t)Q_5^a(t)|0>=0.
\ee 
Hence  we  conclude  that $Q_5^a(t)|\phi_1(t)>=0$.  Next  we  may 
consider  $Q_5^a(t)Q_D(t)|\phi_1(t)>$ and by the  same  reasoning 
obtain the result \be
Q_5^a(t)Q_D(t)|\phi_1(t)>=Q_5^a(t)|\phi_2(t)>= 
Q_D(t)Q_5^a(t)|\phi_1(t)>=0.
\ee
Hence we deduce: $Q_5^a(t)|\phi_2(t)>=0$. In this manner we  find 
that  the axial charge annihilates all the  states  $|\phi_1(t)>, 
|\phi_2(t)>, |\phi_3(t)>$ and so on, {\it ad infinitum\/},  where 
$|0>  \not= |\phi_1(t)> \not= |\phi_2(t)> \not= |\phi_3(t)>$.  We 
thus have deduced the result
\be
Q_5^a(t)|\phi_n(t)>=0, \;\; n=1,2,3, \cdots .
\label{eq19}
\ee

Clearly  Eq.(\ref{eq19})  is satisfied only for  the  Wigner-Weyl 
mode,  i.e., if $|\phi_n(t)>=|0>$, for all $n$, up to  a  unitary 
transformation  $U$  which  commutes with  $\g_5$.  However,  the 
states $|\phi_n(t)>$ are neither vacuum eigenstates nor are  they 
necessarily degenerate since ${\dot Q}_D(t) \not=0$ by virtue  of 
the trace anomaly. As a consequence, any identification of broken 
scale vacuum with chirally symmetric vacuum is wrong and must  be 
rejected. Hence we conclude by {\it reductio ad absurdum\/}  that 
$Q_5^a(t)|0> \not=0$: that chiral symmetry must be  spontaneously 
broken.

We may also infer the connection between the states $|\phi_n(t)>$ 
and  the degenerate vacuum states as follows. Since  $Q_5^a(t)|0> 
\not=0$  and  $Q_D(t)|0>\not=0$, we derive  from  the  commutator 
property    $[Q_D(t),   \;   Q_5^a(t)]   =0$    that    $Q_5^a(t) 
|\phi_1(t)>=Q_D(t)  Q_5^a(t)|0>  = Q_D(t)|0'>$  where  $|0'>$  is 
defined  by $Q_5^a(t)|0>=|0'>\not= |0>$. Hence we determine  that 
$|\phi_1(t)>=(Q_5^a)^{-1}(t)Q_D(t)|0'>$ is the desired connection 
between hidden scale symmetry and hidden chiral symmetry. We note 
that the inverse of $Q_5^a(t)$ exists since $Q_5^a(t)|0>\not=0$.

Spontaneously broken chiral symmetry is sig\-nified by 
the non-va\-nishing anti-com\-mutator in QCD:
\be
\{\; \g_5, S^{-1}(p)\;  \} \not=0.
\ee
The   form   of  the  inverse  quark  propagator  is   given   by 
[\ref{Pagels}] in the limit of zero mass $u,d,s$ quarks:
\be
S^{-1}(p)=\dirac p A(p^2)  - \S_D(p^2)
\ee 
where  $\S_D(p^2)$ is the dynamical quark mass. This  corresponds 
to the form of the quark propagator in the covariant gauge in the 
presence of gluons in Quantum Chromodynamics. 
The  function  $A$  is really a  function  of  the  dimensionless 
variable  $p^2/  \m^2$ where $\m$ is the subtraction  point.  The 
above  arguments  are valid also for $SU(2)  \times  SU(2)$  with 
$f_{abc}$ replaced by $\epsilon_{abc}$.

We can now address the question of Goldstone theorem  [\ref{GSW}] 
in  QCD. Following Cornwall's analysis  [\ref{Cornwallappendix}], 
we  examine  Ward-Takahashi  identity satisfied  by  the  proper, 
renormalized,  color  singlet,  flavor  non-singlet  axial-vector 
current vertex function in QCD in the limit of zero mass  $u,d,s$ 
quarks:
\be
q^{\m}\G^a_{\m 5}(p,p')= Z_A Z_2^{-1}\left [ S^{-1}(p')\g_5 \half 
\l^a + \g_5 \half \l^a S^{-1}(p) \right ]  ,
\ee
where  $q=p-p',  \l^a$ is the flavor $SU(3)$ matrix  and  $a=1,2, 
\cdots,  8$. Since we have established that chiral symmetry  must 
be broken spontaneously, signified by
\be
\left \{ \g_5, S^{-1}(p)\right \} \not= 0,
\ee
it follows that 
\be
\lim_{q  \rightarrow 0} q^{\m}\G^a_{\m 5}(p,p') = -2Z_A  Z_2^{-1} 
\g_5 \half \lambda^a \S_D(p^2)  \not= 0.
\ee
We therefore have
\be
\lim_{q\rightarrow  0}\G_{\m5}^a= \half \lambda^a \left \{-2  Z_A 
Z_2^{-1}\g_5 \S_D(p^2) \frac{q_{\m}}{q^2} + \cdots \right \},
\ee
where  the omitted terms are regular [\ref{Cornwallappendix}]  at 
$q=0$
and  hence the axial-vector vertex function must have a  pole  at 
$q^2 = 0$. This establishes Goldstone theorem in QCD.

In  conclusion,  we  have  dealt  with  the  question  of  chiral 
symmetric  mode  in QCD and the  Nambu-Goldstone  realization  of 
chiral symmetry. It is known that the choice between these two is 
no  longer  free for us. Indeed, ``the theory itself  should,  in 
principle,  tell  us  whether  the symmetry  or  part  of  it  is 
spontaneously broken or not" [\ref{Pokorski}]. This is  precisely  
what we have attempted to do in this work by utilizing  Coleman's 
theorem  and  by  employing the twin  ingredients  of  asymptotic 
freedom and confinement in QCD.

It is interesting to note that Marciano and Pagels [\ref{Marciano}] observed that ``the 
deeper connection between the trace anomaly and the QCD mass spectrum and 
confinement, if it exists, remains to be understood". We believe we have 
shed some light on this question.

\vspace{.2in}
\large
\noindent {\bf REFERENCES AND FOOTNOTES}
\vspace{.2in}

\normalsize 

\begin{enumerate}

\item  \label{CW}  S.Coleman and E.Witten, Phys.  Rev.Lett.  {\bf 
45}, 100 (1980).

\item  \label{tHooft} G.'t Hooft, in {\it Recent Developments  in 
Gauge theories\/}, editors G.t'Hooft {\it et al\/}, Plenum Press, 
(1980)  New  York,  reprinted in {\it  Dynamical  Gauge  Symmetry 
Breaking\/},  edited  by E.Farhi and R.Jackiw,  World  Scientific 
Publishing  Co.  (1982) Singapore ; T.Appelquist,  J.Terning  and 
L.Wijewardhana,  Phys.  Rev.  Lett. {\bf  77},  1214  (1996);  J. 
Cornwall,  Phys.  Rev. {\bf D22}, 1452 (1980). A.  Casher,  Phys. 
Lett.  {\bf  83B}, 395 (1979); O. Nachtman  and  W.Wetzel,  Phys. 
Lett.  {\bf  81B}, 211 (1979); D.Gross and A. Neveu,  Phys.  Rev. 
{\bf  D10},  3235  (1974); K. Lane, Phys. Rev.  {\bf  D10},  1353 
(1974). See also R. Acharya and P. Narayana Swamy, in {\it  A.I.P 
Conference \\
Proceedings  \/} {\bf 150}, ed., D. Geesaman, Lake  Louise  1986, 
American Institute of Physics (1986) New York; C.Callan, R.Dashen 
and D.Gross, Phys. Rev. {\bf D17}, 2717 (1978).

\item  \label{QFT}  S.  Weinberg,  {\it  The  Quantum  Theory  of 
Fields\/},   Volume  II  (1996),   Cambridge  University   Press, 
Cambridge.  See also S.Weinberg, Phys. Rev. Lett. {\bf 65},  1177 
(1990).

\item  \label{AdlerRMP} S. Adler, Rev. Mod. Phys. {\bf  54},  729 
(1982).

\item \label{BJW} K. Johnson, M. Baker and R. Willey, Phys.  Rev. 
{\bf 136}, 1111B (1964).

\item  \label{Coleman}  S.Coleman, J. Math. Phys.  {\bf  7},  787 
(1966). 
\item  \label{PCT} R. Streater and A.Wightman, {\it PCT, Spin  \& 
Statistics, and all that\/}, W. Benjamin, Inc. (1964) New York.

\item  \label{Federbush} P. Federbush and K. Johnson, Phys.  Rev. 
{\bf 120}, 1926 (1960).

\item  \label{Strocchi}  F. Strocchi, Phys. Rev. {\bf  D6},  1193 
(1972).  This  work  extends  the  Federbush-Johnson  theorem  to 
theories  with  indefinite  metric.  Local  gauge  quantum  field 
theories require an indefinite metric. See F. Strocchi, Phys.Rev. 
{\bf D17}, 2010 (1978). 
\item  \label{Collinsetal}  S. Adler, J. Collins and  A.  Duncan, 
Phys.  Rev.  {\bf D15}, 1712 (1977); J. Collins,  A.  Duncan,  S. 
Joglekar, Phys. Rev. {D16}, 438 (1977).

\item \label{Yndurian}  See S. Pokorski, section 6.4, {\it  Gauge 
field theories \/}, Cambridge University Press (1987) Cambridge. 

\item  \label{Zin} J. Zinn-Justin, {\it Quantum field theory  and 
critical  phenomena\/},  second edition,  Clarendon  Press(1993), 
Oxford.

\item \label{Marciano} W. Marciano and H. Pagels, Phys. Rep. {\bf 
36C}, 137 (1978).

\item   \label{Callan}  C.  Callan  in  {\it  Summer  School   of 
Theoretical  Physics, Les Houches 1971, \/} edited by  C.  Dewitt 
and C. Itzykson (1973), Gordon and Breach Publishers.

\item  \label{Pagels}  H.  Pagels, Phys.  Rev.  {\bf  D19},  3080 
(1979).

\item  \label{GSW} J. Goldstone, A. Salam and S. Weinberg,  Phys. 
Rev. {\bf 127}, 965 (1962). See also F.Strocchi, {\it Comm. Math. 
Phys.\/} {\bf 56}, 57 (1977).

\item  \label{Cornwallappendix} See Appendix, J. Cornwall,  Phys. 
Rev. {\bf D22}, 1452 (1980)

{\bf D26}, 

\item  \label{Pokorski} See S.Pokorski, section 9.2,  {\it  Gauge 
field theories \/}, Cambridge University Press (1987) Cambridge. 

\end{enumerate}

\end{document}